\documentclass[aps, 12pt]{revtex4}
\usepackage{graphicx,subfigure}
\usepackage{epsfig}

\usepackage{bbold}
\begin{document}
\title{Steiljes electrostatic model with imaginary charges resolves wave-particle duality }
\author{K V S Shiv Chaitanya}
\email[]{ chaitanya@hyderabad.bits-pilani.ac.in}
\affiliation{Department of Physics,, BITS Pilani, Hyderabad Campus, Jawahar Nagar, \\Shamirpet Mandal,
Hyderabad, India 500 078.}
\author{V Srinivasan}
\email[]{ vsspster@gmail.com}
\affiliation{Department of Theoretical Physics, University of Madras,  Guindy, Chennai, India, 600025.}

\begin{abstract}
In this paper, we show that the quantum bound state problems are mapped to  $N$ point vortices with the  identical  circulation or strength using the  Steiljes electrostatic model  with imaginary charges.  We also show that  these $N$ point charges or vortices, become imaginary, in a constant background field will admit a wave solution under paraxial wave approximation. Therefore, in quantum mechanics as long as paraxial approximation is valid it behaves like a wave.
\end{abstract}

\maketitle

\section{Introduction}

One of the major point of difference between the classical mechanics and quantum mechanics is the Heisenberg uncertainty relationship. This is nothing but the restatement of the wave particle duality. The wave particle duality is the corner stone of  quantum mechanics, it states that quantum particles are both particles and wave. There are several theories based on this wave particle duality of which the most widely accepted is the principle of complementarity. The essence of this  principle is that quantum systems exhibit both the wave and the particle nature, but,  the outcome of an event is purely depends on a given experiment.

In this paper, we would like to address the wave particle duality using the paraxial wave optics. We show that as long as paraxial wave approximation is valid in a given experimental situation the quantum system behaves like a wave. The best example is the double slit experiment, in this experiment, the electrons pass either through any one of the two slits and strikes the screen the one find a diffraction pattern. The input for the experiment are the electrons which are particles and the output detected is a wave. This is wave particle duality. In the language of optics, the double slit comes under the scalar diffraction theory,  the Fresnel diffraction integral is derived using the Fresnel approximation which is nothing but the paraxial approximation \cite{born}. For the far field the integral reduces to a Fourier transform, then the convolution of two delta functions in place of  the double slits with two gives rise to interference. Hence, in the double slit experiment of electrons the paraxial wave approximation is valid.

To show this, we use the quantum Hamilton Jacobi formalism. The quantum Hamilton Jacobi can mapped to the Steiljes electrostatic model  with imaginary charges \cite{kvs}. This Steiljes electrostatic model   with imaginary charges  can also be mapped to $N$ point vortices with the  identical  circulation or strength of all the vortices \cite{aref}. Once the state of the quantum system are associated with the imaginary charges it is  shown that these charges or vortices  with a constant background field under paraxial wave approximation will admit a wave solution. Therefore, in quantum mechanics as long as paraxial wave approximation is valid it behaves like a wave. The application of paraxial optics to quantum harmonic oscillators is studied in ref \cite{prh}.

 In our model, the wave function  is related to quantum momentum function by a pure phase. This  quantum momentum function has singularities which are complex in nature, we identify these singularity with  imaginary charges or  imaginary vortices. This singularities are the moving singularities.This model has close resemblance to the pilot wave theory of double solution \cite{de}. The  pilot wave theory was initially proposed by de Broglie as the theory of double solution. In this theory, the Schr\"odinger equation satisfies two solutions one the regular wave function $\psi(x,y,z)=ae^{i\phi(x,y,z)}$ where $a$ is a constant and the another a physical wave solution $u(x,y,z)=f(x,y,z)e^{i\phi(x,y,z)}$ and the two solutions are related by the phase $\phi(x,y,z)$. The singularities in $u(x,y,z)$ behaves like a particle. This singularities are the moving singularities. The draw back of the theory is that it explains only for the single particle case. This theory was further developed by Bohm \cite{bo,bo1} in which the many particle case is developed. In this theory, the wave particle duality vanishes, a quantum system behaves like particle  and the wave simultaneously then the particles are directed by the pilot wave which will guide them to areas of interference. 
 
 We draw the following analogy with pilot wave theory and our model, the singularities in both the model are moving singularities. In pilot wave theory they are associated with particles and in our model these singularities are complex in nature and we associated with the imaginary charges. The difference between our model and the pilot wave theory is that we have one solution and they have two solutions.
 
 The paper is organised as follows, in section II we describe the connection between the quantum Hamilton Jacobi and the Steiljes electrostatic model  with imaginary charges followed by $N$ Point vortices description in the section III then followed is our main result of the paper in section IV and finally conclusion in section V.

\section{Quantum Hamiltonian Jacobi Formalism}
The Schr\"odinger equation is given by,
\begin{equation}
- \frac{\hbar^2}{2m}\frac{\partial^2 }{\partial x^2} \psi(x) + V(x) \psi(x) = E
  \psi(x).    \label{sc} 
\end{equation}
In quantum Hamilton Jacobi formalism, one defines a function $S$ analogous to the classical characteristic function by the relation 
\begin{equation}
\psi(x) = \exp\left(\frac{iS}{\hbar}\right)       \label{ac}
\end{equation}which, when substituted in (\ref{sc}), gives
\begin{equation}
p^2 - i \hbar \frac{dp}{dx} = 2m (E - V(x)),       \label{qhj1}
\end{equation}
where quantum momentum function $p=\frac{\partial S}{\partial x}$ and the differential equation known as the Riccati equation. then the relation between the wave function $\psi(x)$ and quantum momentum function is given by 
\begin{equation}
p = -i \hbar\frac{\partial ln \psi(x)}{\partial x}  \label{lg}
\end{equation}

 The complex pole expansion of quantum Hamilton Jacobi equation (\ref{qhj1}), for the operator  $p$ the quantum momentum function,  in terms of $n$ moving poles and fixed poles \cite{bhal, bhal1, sree, geo1},  is given by 
\begin{equation}
p(x)=\sum_{k=1}^n\frac{-i\hbar}{x-x_k}+Q(x).\label{ufo}
\end{equation}
Its is proved in ref\cite{cho}, the complex poles expansions of the form 
of nonlinear partial differential equations or Burger's equation is satisfied if and only if  the following system of differential equations are satisfied or the complex poles evolve according to the system of $n$ linear equation given by
\begin{equation}
\frac{d x_j}{dt}=\sum_{1\leq j\leq n,j\neq k}^n\frac{1}{x_k-x_j}+Q(x_j),\label{ufie}
\end{equation}
 in each equation $j=k$  term is not present and there are no poles in $Q(x_j)$ at $x_j$. The relation between the Riccati equation and the Burger equation is given in ref \cite{R}.  In the equation (\ref{ufo}) there are $n$ poles and the equation (\ref{ufie}) there are $n$ equations corresponding to $n$ poles
\begin{equation}
\sum_{1\leq j\leq n,j\neq k}^n\frac{-i\hbar}{x_k-x_j}+Q(x_j)=0.\label{ufiu}
\end{equation}
This equation is identical to the Steiljes electrostatic model \cite{st,st1}. 
The solution of system of $n$ linear equation is solved by using the following Identity  \cite{met}
\begin{eqnarray}
\sum_{k=1}\frac{1}{x-x_k}=\frac{f'(x)}{f(x)}-\frac{1}{x-x_j}=\frac{(x-x_j)f'(x)-f(x)}{(x-x_j)f(x)},\label{dig}
\end{eqnarray}
where
\begin{eqnarray}
f(x)=(x-x_1)(x-x_2)\cdots (x-x_n),\label{poly}
\end{eqnarray}
by taking the following limit $x\rightarrow x_j$ and using l'Hospital rule one gets
\begin{eqnarray}
\sum_{1\leq j\leq n,j\neq k} \frac{1}{x_j-x_k}&=&\lim_{x\rightarrow x_j}
\left[\frac{f'(x)}{f(x)}-\frac{1}{x-x_j}\right]\nonumber\\&=&\lim_{x\rightarrow x_j}\frac{(x-x_j)f'(x)-f(x)}{(x-x_j)f(x)}
\nonumber \\&=& \frac{f''(x_j)}{2f'(x_j)}.\label{id}
\end{eqnarray}
By substituting the equation (\ref{id}) in equation (\ref{ufiu}) reduces to 
\begin{equation}
-i\hbar \frac{f''(x_j)}{2f'(x_j)}+Q(x_j)=0.\label{pok}
\end{equation}

In the Steiljes electrostatic model the $\sum_{1\leq j\leq n,j\neq k} \frac{1}{x_j-x_k}$ represents the moving charges between the two fixed charges $Q(x_j)$.  It is shown by author himself that the 
quantum Hamilton Jacobi formalism is analogous to the Steiljes electrostatic model as shown in ref \cite{kvs} and all the bound state problems of quantum mechanics can be solved using Steiljes electrostatic model with identifying the quantum states with imaginary charges.
In this paper we show the  equation (\ref{ufiu}) is exactly identical to the stationary point vortices with identical circulation or strength $\Gamma=-i\hbar$ at positions $x_1\cdots x_N$  with a background flow $Q(x_j)$. 

\section{Point Vortices}
The classical mechanics of $N$ point vortices is described on the unbounded xy plane
with circulation or strength  $\Gamma_\alpha$  at the positions $(x_\alpha, y_\alpha)$, where
 $\alpha=1\cdots N$. by
$2N$ first-order, nonlinear, ordinary differential equations:

\begin{equation}
\frac{d x_\alpha}{dt}=\sum_{\alpha\neq\beta,\beta=1}^n\frac{\Gamma_\beta(x_\alpha-x_\beta)}{l_{\alpha \beta}^2},\;\;\;
\frac{d y_\alpha}{dt}=\sum_{\alpha\neq\beta,\beta=1}^n\frac{\Gamma_\beta(y_\alpha-y_\beta)}{l_{\alpha \beta}^2}\label{pt}
\end{equation}
where $l_{\alpha \beta}^2=(x_\alpha-x_\beta)^2+(y_\alpha-y_\beta)^2$.  The $2N$ first-order, nonlinear, ordinary differential equations are reduced to $N$ first-order, nonlinear, ordinary differential equations by going to a complex plane. Then the position of $N$ vortices are given by   $z_\alpha=x_\alpha+iy_\alpha$ with circulation or strength  $\Gamma_\alpha$ and equation of motion is given by
\begin{equation}
\frac{d \bar{z}_\alpha}{dt}=\sum_{1\leq \alpha\leq n,\alpha\neq \beta}^n\frac{\Gamma_\beta}{z_\alpha-z_\beta},\label{pt1}
\end{equation}
where the overline denotes complex conjugation. 
Kirchhoff has given the Hamiltonian formalism for the $N$ point vortex model and the Poison bracket 
 in terms of the complex coordinates given,
\begin{equation}
[z_\alpha,z_\beta]=0,\;\; [z_\alpha,\bar{z}_\beta]=-2i\delta_{\alpha\beta}\Gamma_\alpha.\label{com}
\end{equation}
For more details refer to \cite{aref}.

\section{Wave particle duality}
The two-dimensional model of point vortices in an incomprehensible fluid  with $N$ point vortices with circulation or strength  $\Gamma_1\cdots \Gamma_N$ at positions $z_1\cdots z_N$ with background flow $W(z_\beta)$ equation of motion is governed  by the Helmholtz’s equations (\ref{pt1}), in the stationary case $\frac{d \bar{z}_\alpha}{dt}=0$, hence
\begin{equation}
\sum_{1\leq \alpha\leq n,\alpha\neq \beta}^n\frac{\Gamma_\beta}{z_\alpha-z_\beta}+ W(z_\beta)=0.\label{pt11}
\end{equation}
Theorem: \textit{The  two dimensional $N$ point vortices model with the identical circulation strength reduces to the Steiljes electrostatic model}\cite{aref}.

Proof: The Helmholtz’s equations (\ref{pt11}) for two dimensional point vortices model with the identical circulation strength reduces to
\begin{equation}
\sum_{1\leq \alpha\leq n,\alpha\neq \beta}^n\frac{\Gamma}{z_\alpha-z_\beta}+ W(z_\beta)=0,\label{p11}
\end{equation}
Substituting the identity in equation (\ref{id}) in equation (\ref{p11}) reduces to
\begin{equation}
\Gamma \frac{f''(z_j)}{2f'(z_j)}+ W(z_\beta)=0.\label{po}
\end{equation}
The equation (\ref{po}) can be solved using Steiljes electrostatic model. Thus, the equation (\ref{p11}) is identical to equation (\ref{ufiu}) provided circulation or strength $\Gamma=-i\hbar$ are same at positions $x_1\cdots x_N$  with a background flow $W(z_\beta)$.
Hence, the background flow $W(z_\beta)$ is such that it  represents the fixed charges in the Steiljes electrostatic model for point vortices.   Therefore, the system of point vortices has equilibrium at the zeroes of the classical orthogonal polynomials depending on the position of the fixed charges for details refer to \cite{st,st1}.
To address the wave particle duality we state the following theorem.

Theorem: \textit{The  two dimensional $N$ point vortices with the identical imaginary circulation strength under the constant background flow will admit a wave equation and under paraxial approximation}. 

Proof: We will start the equation (\ref{p11}) with all the vortices have equal circulation or strength $\Gamma$ and under constant background field $W(z_\beta)=P$ is given as
\begin{equation}
\sum_{1\leq \alpha\leq n,\alpha\neq \beta}^n\frac{\Gamma}{z_\alpha-z_\beta}+ P=0.\label{pt2}
\end{equation}
Substituting the identity in equation (\ref{id}) in equation (\ref{pt2}) reduces to
\begin{equation}
\frac{\partial^2 u}{\partial z_i^2}+2k\frac{\partial u}{\partial z_i}=0\label{hz4}
\end{equation}
where $k=P/\Gamma$. Putting back the the position of $N$ vortices  $z_\alpha=x_\alpha+iy_\alpha$  in the  equation (\ref{hz4}) reduces to
\begin{equation}
\frac{\partial^2 u}{\partial x_i^2}+\frac{\partial^2 u}{\partial y_i^2}+2k\frac{\partial u}{\partial z_i}=0.\label{hz3}
\end{equation}
If the charges are vortices become imaginary that is $\Gamma=-i\hbar$ the differential equation (\ref{hz3}) becomes
\begin{equation}
\frac{\partial^2 u}{\partial x_i^2}+\frac{\partial^2 u}{\partial y_i^2}+2ik\frac{\partial u}{\partial z_i}=0.\label{hz2}
\end{equation}
In optics this wave equation is called paraxial wave equation. In the paraxial approximation they neglect $\frac{\partial^2 u}{\partial z_i^2}$ term putting back this term in the  equation (\ref{hz2}) reduces to
\begin{equation}
\frac{\partial^2 u}{\partial x_i^2}+\frac{\partial^2 u}{\partial y_i^2}+\frac{\partial^2 u}{\partial z_i^2}
+2ik\frac{\partial u}{\partial z_i}=0,\label{hz1}
\end{equation}
this equation is derived from
the Helmholtz equation:
\begin{equation}
\nabla^2\psi(x_i,y_i,z_i)+k^2\psi(x_i,y_i,z_i)=0\label{hz}
\end{equation} 
with
\begin{equation}
\psi(x_i,y_i,z)=u(x_i,y_i,z_i)e^{ikz_i}.\label{so}
\end{equation}
From this  it is clear that the imaginary point vortices or charges under constant background flow and with the  paraxial approximation satisfy wave equation. Therefore, the point vortices or charges in Steiljes electrostatic model becoming complex gives rise to wave-particle duality. The wave particle duality is associated with the Heisenberg uncertainty relation. We claim that their exist a the Heisenberg uncertainty relation in terms of poison brackets

We also point out that the Poison bracket equation
 $[z_\alpha,\bar{z}_\beta]=-2i\delta_{\alpha\beta}\Gamma_\alpha$ or $[Q,P]=\sum_{\alpha=1}^N\Gamma_\alpha$ are  the classical version of Heisenberg uncertainty relation. 
If we consider the Steiljes electrostatic model, in this case the  vortices have the identical strength $\Gamma=-i\hbar$ and the poison brackets reduce to 
$[z_\alpha,\bar{z}_\beta]=2\hbar$ or $[Q,P]=-i\hbar$. The Heisenberg uncertainty relation is associated with wave particle duality thus by making the circulation or strength imaginary point vortices should admit a wave equation.

\section{Conclusion}
In this paper, we have shown that the quantum bound state problems are mapped to  $N$ point vortices with the  identical  circulation or strength using the  Steiljes electrostatic model  with imaginary charges.  We have also shown that  these $N$ point charges or vortices, become imaginary, in a constant background field will admit a wave solution under paraxial wave approximation. Therefore, we conclude that in quantum mechanics as long as paraxial approximation is valid it behaves like a wave.

\section*{Acknowledgments}
Authors thank KVSSC acknowledges the Department of Science and technology, Govt of India (fast track scheme (D. O. No: SR/FTP/PS-139/2012)) for financial support.

\end{document}